\documentclass[article ,aps, prx, amsmath, amssymb]{revtex4-2}
\usepackage{hyperref}
\usepackage{amsfonts}
\usepackage{graphicx}
\usepackage{mathtools}
\usepackage{float}

\begin{document}

\newcommand{\ket}[1]{|#1\rangle}
\newcommand{\bra}[1]{\langle #1 |}
\title{Derivation of the Weizsäcker Density Functional from Probability Theory}
\author{Connor Dolan}
\affiliation{Department of Physics, University at Albany, State University of New York, Albany, New York 12222, USA}
\date{\today}

\begin{abstract}
We demonstrate that the Weizsäcker potential is an exact term in the universal functional in density functional theory (DFT) for the ground state 
of a system with $N$ electrons. This proof uses no approximations or physical arguments, and follows from the form of kinetic energy of the ground state
and probability theory. We also examine the form of the other terms in the kinetic energy. 
\end{abstract}
\maketitle

\section{Universal Functional}

The Hohenberg-Kohn theorems state that the ground state of $N$ electrons in a potential is determined uniquely by the electron density \cite{Hohenberg1964} . 
This implies the existence of a universal functional that accounts for the kinetic energy and electron-electron potential energies.

\begin{equation}
F_N[n]=T_N[n]+V_{ee, N}[n]
\end{equation}

Where $T_N[n]$ is the kinetic energy and $V_{ee, N}[n]$ is the electron-electron interaction energy. \\

The existence of the universal functional implies that any term in $F_N[n]$ found using first principles must 
be an exact term in the the universal functional.\\

\section{Kinetic Energy}

The wavefunction $\Psi$ for a system on $N$ electrons is given by:

\begin{equation}
\Psi(r_1,..., r_N)=\sqrt{\rho(r_1,..., r_N)}e^{i\phi(r_1,..., r_N)}
\end{equation}

Where $\rho$ is the joint probability, $\phi$ is the phase, and $r_n$ is the position of the $n$-th electron. \\

The kinetic energy is given by \cite{sakurai_modern_1994}:

\begin{equation}
\begin{split}
T&=\int d^3r_1 ...\int d^3 r_N \Psi^*(r_1,...,r_N)\left(-\frac{1}{2}\sum_i \nabla^2_i \right) \Psi(r_1,...,r_N)\\
&=-\frac{N}{2}\int d^3r_1 ...\int d^3 r_N \Psi^*(r_1,...,r_N)\nabla^2_1\Psi(r_1,...,r_N)\\
&=\frac{N}{2}\int d^3r_1 ...\int d^3 r_N \rho(r_1,...,r_N)
\left({\left|\frac{1}{2}\nabla_1\log(\rho(r_1,...,r_N))\right|}^2+{\left|\nabla_1\phi(r_1,...,r_N)\right|}^2\right) \\
\end{split}
\end{equation}

Where we have used the fact that the wavefunction is antisymmetric to change variables for the Laplacian of the position of each electron to operate only
on $r_1$ N times.\\

No other terms in the electron-electron interactions or the potential depend on the phase $\phi$. If we wish to find the ground state, 
we minimize the energy, and the minimum of the phase dependent term in our kinetic energy is zero and occurs when the phase is a constant. \\

The kinetic energy for the ground state is then:

\begin{equation}
T=\frac{N}{8}\int d^3r_1 ...\int d^3 r_N \rho(r_1,...,r_N){\left| \nabla_1\log(\rho(r_1,...,r_N))\right|}^2
\end{equation}

\section{The Weizsäcker Functional}

The joint probability distribution $\rho(r_1,...,r_N)$ can be rewritten as a single electron probability distribution and a conditional probability:

\begin{equation}
\rho(r_1,...,r_N)=\rho(r_2,...,r_N|r_1)\rho(r_1)
\end{equation}

This gives us: 
\begin{equation}
\begin{split}
&\log(\rho(r_1,...,r_N))=\log(\rho(r_2,...,r_N|r_1))+\log(\rho(r_1))\\
\implies & \nabla_1\log(\rho(r_1,...,r_N))=\nabla_1\log(\rho(r_2,...,r_N|r_1))+\nabla_1\log(\rho(r_1))\\
\end{split}
\end{equation}

The kinetic energy is then:

\begin{equation}\label{eq:kinetic}
\begin{split}
T&=\frac{N}{8}\int d^3r_1 ...\int d^3 r_N \rho(r_1,...,r_N)\Bigl(
{\left| \nabla_1\log(\rho(r_1))\right|}^2\\
&+2\nabla_1 \log(\rho(r_1))\cdot\nabla_1\log(\rho(r_2,...,r_N|r_1))+
{\left| \nabla_1\log(\rho(r_2,...,r_N| r_1))\right|}^2 \Bigr)\\
\end{split}
\end{equation}

Let us denote the three terms in the order they are shown as $T_1$, $T_2$ and $T_3$ so that 
the kinetic energy is $T=T_1+T_2+T_3$. The first term can be simplified:

\begin{equation}
\begin{split}
T_1&=\frac{N}{8}\int d^3r_1 ...\int d^3 r_N \rho(r_1,...,r_N)\left| \nabla_1\log(\rho(r_1))\right|^2 \\
&=\frac{N}{8}\int d^3r_1 ...\int d^3 r_N \rho(r_2,...,r_N|r_1)\rho(r_1)\left| \nabla_1\log(\rho(r_1))\right|^2 \\
&=\frac{N}{8}\int d^3r_1\rho(r_1)\left| \nabla_1\log(\rho(r_1))\right|^2 \\
&=\frac{N}{8}\int d^3r\rho(r)\left| \nabla\log(\rho(r))\right|^2 \\
&=\frac{N}{8}\int d^3r\frac{\left| \nabla\rho(r)\right|^2}{\rho(r)} \\
\end{split}
\end{equation}

Where the conditional probability $\rho(r_2,...,r_N| r_1)$ factors out  and integrates to $1$, and we relabel $r_1$ as $r$. \\

The density is equal to the number of electrons times the probability distribution of one electron, $n(r)=N\rho(r)$, so we obtain 
the kinetic energy term:

\begin{equation}
T_1=\frac{1}{8}\int d^3r\frac{\left| \nabla n(r)\right|^2}{n(r)} \\
\end{equation}

Which is the Weizsäcker functional\cite{weizsacker_density_1935}. \\ 

This means the Weizsäcker functional is an exact term in the universal density functional. 

\newpage 

\section{Other Kinetic Energy Terms}

We now examine the other terms in (\ref{eq:kinetic}). 
The second term $T_2$:

\begin{equation}
\begin{split}
T_2&=\frac{N}{4}\int d^3r_1 ...\int d^3 r_N \rho(r_1,...,r_N) \nabla_1 \log(\rho(r_1))\cdot\nabla_1\log(\rho(r_2,...,r_N|r_1))\\
&=\frac{N}{4}\int d^3r_1 ...\int d^3 r_N \rho(r_2,...,r_N|r_1) \nabla_1 \log(\rho(r_1))\cdot\nabla_1\log(\rho(r_2,...,r_N|r_1))\\
&=\frac{N}{4}\int d^3r_1 ...\int d^3 r_N \nabla_1 \rho(r_1)\cdot\nabla_1 \rho(r_2,...,r_N|r_1)\\
&=\frac{N}{4}\int d^3r_1 \nabla_1 \rho(r_1)\cdot\nabla_1\left(\int d^3 r_2 ... \int d^3 r_N  \rho(r_2,...,r_N|r_1) \right)\\
&=\frac{N}{4}\int d^3r_1 \nabla_1 \rho(r_1)\cdot\nabla_1 1=0\\
\end{split}
\end{equation}

$T_2$ vanishes. $T_3$ gives us:

\begin{equation}
\begin{split}
T_3&=\frac{N}{8}\int d^3r_1 ...\int d^3 r_N \rho(r_1,...,r_N){\left| \nabla_1\log(\rho(r_2,...,r_N| r_1))\right|}^2\\
&=\frac{N}{8}\int d^3r_1 ...\int d^3 r_N \rho(r_2,...,r_N|r_1)\rho(r_1){\left| \nabla_1\log(\rho(r_2,...,r_N| r_1))\right|}^2\\
&=\int d^3r_1 \rho(r_1)T(r_1)=\int d^3r \rho(r)T(r)\\
\end{split}
\end{equation}

Where we relabeled $r_1$ to $r$ and defined a kinetic energy function given by:

\begin{equation}
T(r)=\frac{N}{8}\int d^3r_2 ...\int d^3 r_N \rho(r_2,...,r_N|r){\left| \nabla\log(\rho(r_2,...,r_N|r))\right|}^2\\
\end{equation}

The information metric \cite{caticha_entropic_2006} for the conditional probability distribution of all but one electron positions on that of the position of one electron, 
is given by:

\begin{equation}
g_{ij}(r) = \int d^3r_2...\int d^3 r_N \rho(r_2,...,r_N|r)\nabla_i\log(
\rho(r_2,...,r_N|r))\nabla_j\log(\rho(r_2,...,r_N|r))
\end{equation}

We then see that the kinetic energy function $T(r)$ is proportional to the trace of the information metric:

\begin{equation}
T(r) = \frac{N}{8} \sum_i g_{ii}(r)
\end{equation}

So that $T_3$ is given as:

\begin{equation}
T_3 = \frac{N}{8} \int d^3 r \rho(r) \sum_i g_{ii}(r) = \frac{1}{8}\int d^3 r n(r) \sum_i g_{ii}(r)
\end{equation}

This result substantiates previous work that finds the kinetic energy is proportional to the information metric \cite{IpekCaticha2016}, 
and explorations of the relationship between Density Functional Theory and information physics more generally \cite{Yousefi2021}.

\bibliography{references}

\end{document}